\documentclass[prd,aps,a4,twocolumn,superscriptaddress,preprintnumbers,nofootinbib]{revtex4-1}
\usepackage{pslatex}
\usepackage[pdftex]{graphicx}
\usepackage{psfrag}
\usepackage{epsfig}
\usepackage{color}
\usepackage{cancel}
\usepackage{slashed}
\usepackage{amssymb}
\usepackage{amsmath}
\usepackage{hyperref}
\usepackage{enumerate}
\usepackage{multirow}
\usepackage{natbib}
\usepackage[normalem]{ulem}
\begin{document}
\title{CP violating effects in coherent elastic
  neutrino-nucleus scattering processes}%
\author{D. Aristizabal Sierra}%
\email{daristizabal@ulg.ac.be}%
\affiliation{Universidad T\'ecnica
  Federico Santa Mar\'{i}a - Departamento de F\'{i}sica\\
  Casilla 110-V, Avda. Espa\~na 1680, Valpara\'{i}so, Chile}%
\affiliation{IFPA, Dep. AGO, Universit\'e de Li\`ege, Bat B5, Sart
  Tilman B-4000 Li\`ege 1, Belgium}%
\author{V. De Romeri}%
\email{deromeri@ific.uv.es}%
\affiliation{AHEP Group, Instituto de F\'{i}sica Corpuscular,
  CSIC/Universitat de Val\`encia,\\ Calle Catedr\'atico Jos\'e
  Beltr\'an, 2 E-46980 Paterna, Spain}%
\author{N. Rojas}%
\email{nicolas.rojasro@usm.cl}%
\affiliation{Universidad T\'ecnica
  Federico Santa Mar\'{i}a - Departamento de F\'{i}sica\\
  Casilla 110-V, Avda. Espa\~na 1680, Valpara\'{i}so, Chile}%
\begin{abstract}
  The presence of new neutrino-quark interactions can enhance, deplete
  or distort the coherent elastic neutrino-nucleus scattering
  (CE$\nu$NS) event rate. The new interactions may involve CP
  violating phases that can potentially affect these
  features. Assuming light vector mediators, we study the effects of
  CP violation on the CE$\nu$NS process in the COHERENT sodium-iodine,
  liquid argon and germanium detectors. We identify a region in
  parameter space for which the event rate always involves a dip and
  another one for which this is never the case. We show that the
  presence of a dip in the event rate spectrum can be used to
  constraint CP violating effects, in such a way that the larger the
  detector volume the tighter the constraints. Furthermore, it allows
  the reconstruction of the effective coupling responsible for the
  signal with an uncertainty determined by recoil energy
  resolution. In the region where no dip is present, we find that CP
  violating parameters can mimic the Standard Model CE$\nu$NS
  prediction or spectra induced by real parameters. We point out that
  the interpretation of CE$\nu$NS data in terms of a light vector
  mediator should take into account possible CP violating
  effects. Finally, we stress that our results are qualitatively
  applicable for CE$\nu$NS induced by solar or reactor
  neutrinos. Thus, the CP violating effects discussed here and their
  consequences should be taken into account as well in the analysis of
  data from multi-ton dark matter detectors or experiments such as
  CONUS, $\nu$-cleus or CONNIE.
\end{abstract}
\maketitle
\section{Introduction}
\label{sec:intro}
Coherent elastic neutrino-nucleus scattering (CE$\nu$NS) is a process
that occurs when the de Broglie wavelength $\lambda$ of the scattering
process is larger than the nuclear radius. In terms of the exchanged
momentum $q$ this means that when $q\lesssim h/r_N\simeq 100\,$MeV the
individual nucleonic amplitudes sum up coherently. As a consequence
the total amplitude gets enhanced by the number of nucleons, resulting
in a rather sizable cross section. Indeed, among all possible
scattering processes at neutrino energies below $100\,$MeV, CE$\nu$NS
has the largest cross section. Measuring CE$\nu$NS however is
challenging due to the small nuclear recoil energies involved. The
first measurement was done in 2017 by the COHERENT experiment, which
observed the process at a 6.7$\,\sigma$ confidence level (CL), using
neutrinos produced in the Oak Ridge National Laboratory Spallation
Neutron Source~\cite{Akimov:2017ade}.

Given the constraints on the neutrino energy probe, CE$\nu$NS can be
induced by neutrinos produced in fixed target experiments such as in
COHERENT, reactor neutrinos and solar and atmospheric
neutrinos. Within the second category CONUS is an ongoing experiment
\cite{conus} and there are as well other experimental proposals that
aim at using reactor neutrinos to measure CE$\nu$NS using different
technologies \cite{Aguilar-Arevalo:2016khx,Strauss:2017cuu}. Relevant
for the third category are direct detection multi-ton dark matter (DM)
experiments such as XENONnT, LZ and DARWIN
\cite{Aprile:2015uzo,Akerib:2018lyp,Aalbers:2016jon}. There is clearly
a great deal of experimental interest on CE$\nu$NS, in particular for
the role it will play in near-future DM direct detection experiments
\cite{Billard:2013qya,Dutta:2019oaj} and the different physics
opportunities it offers in these facilities
\cite{Harnik:2012ni,Cerdeno:2016sfi,Shoemaker:2017lzs,Dutta:2017nht,AristizabalSierra:2017joc,Gonzalez-Garcia:2018dep,Billard:2018jnl}. From
the phenomenological point of view, it is therefore crucial to
understand the different uncertainties the process involves and the
impact that new physics effects might have on the predicted spectra.

The Standard Model (SM) CE$\nu$NS cross section proceeds through a
neutral current process \cite{Freedman:1973yd,Freedman:1977xn}
\footnote{Recently this cross section has been revisited and the
  incoherent neutrino-nucleus elastic cross section has been
  recalculated in \cite{Bednyakov:2018mjd}.}. Depending on the target
nucleus, in particular for heavy nuclei, it can involve sizable
uncertainties arising mainly from the root-mean-square radius of the
neutron density distribution
\cite{AristizabalSierra:2019zmy}. However, apart from this nuclear
physics effect the SM provides rather definitive predictions for
CE$\nu$NS on different nuclear targets. Precise measurements of the
process offer a tool that can be used to explore the presence of new
physics effects. In fact, since the COHERENT data
release~\cite{Akimov:2017ade,Akimov:2018vzs}, various analyses
involving new physics have been carried out. The scenarios considered
include effective neutrino non-standard interactions
\cite{Coloma:2017ncl,Liao:2017uzy,Miranda:2019skf}, light vector and
scalar mediators \cite{Liao:2017uzy,Farzan:2018gtr}, neutrino
electromagnetic properties \cite{Kosmas:2017tsq,Miranda:2019wdy},
sterile neutrinos \cite{Kosmas:2017tsq} and neutrino generalized
interactions \cite{AristizabalSierra:2018eqm}.

Analyses of new physics so far have considered CP conserving
physics. This is mainly motivated by simplicity
and---arguably---because at first sight one might think that getting
information on CP violating interactions in CE$\nu$NS experiments is
hard, if possible at all. CP violating effects are typically studied
through observables that depend on asymmetries that involve states and
anti-states or polarized beams, which in a CE$\nu$NS experiment are
challenging to construct. In this paper we show that information on CP
violating interactions can be obtained in a different way through the
features they induce on the event rate spectrum, and for that aim we
consider light vector mediator scenarios (with masses
$m_V\lesssim 100\,$MeV).

Phenomenologically, among the possible new degrees of freedom that can
affect CE$\nu$NS, light vectors are probably the most suitable. In
contrast to heavy vectors, they are readily reconcilable with
constraints from the charged lepton sector, while at the same time
leading to rather sizable effects \cite{Dent:2016wcr}. In contrast to
light scalar mediators, they interfere with the SM contribution and
can eventually lead to a full cancellation of the event rate at a
specific nuclear recoil energy. This is a feature of particular relevance in
the identification of CP violating effects, as we will show.

In our analysis we use the COHERENT germanium (Ge), sodium (Na) and liquid argon (LAr)
detectors to show the dependence of CP violating effects on target materials
and detector volumes. We fix the detector parameters according
to future prospects \cite{scholberg} and in each case we extract
information of CP violation by comparing CP conserving and CP
violating event rate spectra (induced by real or complex
parameters). We then establish the reach of each detector to
constrain CP violating effects by performing a $\chi^2$ analysis.

The rest of the paper is organized as follows. In
sec. \ref{sec:cp-violating-int} we fix the interactions, the notation
and we introduce the parametrization that will be used throughout our
analysis. In sec. \ref{sec:param-space-analysis} we present the
parameter space analysis, we discuss constraints on light vector
mediators and identify CP violating effects. In
sec. \ref{sec:CPV-effects} we discuss the possible limits that the
sodium, germanium and argon detectors could eventually establish on CP
violating effects. Finally, in sec. \ref{sec:conclusions} we summarize
our results.
\section{CP violating interactions}
\label{sec:cp-violating-int}
Our analysis is done assuming that the new physics corresponds to the introduction of
light vector mediators. This choice has to do with phenomenological
constraints. Although subject to quite a few number of limits, models
for such scenarios already exist \cite{Farzan:2015doa}. They are not only
phenomenologically consistent, but they also allow for large effects in a
vast array of experiments \cite{Farzan:2015hkd,Farzan:2016wym}. In
contrast, in heavy mediator models the constraints from the charged
lepton sector lead---in general---to effective couplings whose effects
barely exceed few percent \cite{Wise:2014oea}.

We allow for neutrino vector and axial currents, while for quarks we
only consider vector interactions (axial quark currents are spin
suppressed), and we assume that all
couplings are complex at the renormalizable level. 
The Lagrangian of the new physics can then be
written according to
\begin{align}
  \label{eq:BSM-Lag}
  \mathcal{L}=f_V\overline{\nu}\gamma_\mu\nu\,V^\mu 
  + 
  if_A\overline{\nu}\gamma_\mu\gamma_5\nu\,V^\mu
  +
  \sum_{q=u,d} h^q_V\overline{q}\gamma_\mu q\,V^\mu\ ,
\end{align}
where $f_V=|f_V|e^{i\phi_V}$, $f_A=|f_A|e^{i\phi_A}$,
$h_V^q=|h_V^q|e^{i\phi_{Vq}}$, we have dropped lepton flavor indices
and we restrict the sum to first generation quarks. In terms of the
``fundamental'' parameters the nuclear vector current coupling reads
(with explicit dependence on the transferred momentum $q$)

\begin{equation}
  \label{eq:nuclear-vector-coupling}
  h_V(q^2)=N\left(2h_V^d+h_V^u\right)
    F_n(q^2)
    +
    Z\left(h_V^d+2h_V^u\right)F_p(q^2)\ ,
\end{equation}
where $N=A-Z$, with $A$ and $Z$ the mass and atomic number of the
corresponding nuclide. $F_{n,p}(q^2)$ are the neutron and proton
nuclear form factors obtained from the Fourier transform of the
nucleonic density distributions (in the first Born
approximation). Note that this differentiation is particularly
relevant for nuclides with $N>Z$, such as sodium, argon or germanium
\cite{AristizabalSierra:2019zmy}.

The interactions in (\ref{eq:BSM-Lag}) affect CE$\nu$NS processes, as
they introduce a $q$ dependence, absent in the SM, that changes the
recoil energy spectrum and can either enhance or deplete the expected
number of events. Here we will consider both mono- and multi-target
detectors, and so we write the CE$\nu$NS cross section for the
$i^{\underline{\text{th}}}$ isotope:
\begin{equation}
  \label{eq:x-sec}
  \frac{d\sigma}{dE_r}=\frac{G_F^2m_i}{2\pi}\left|\xi_V(q_i^2)\right|^2
  \left(
    2 
    - \frac{E_rm_i}{E_\nu^2} 
    - \frac{2E_r}{E_\nu}
    + \frac{E_r^2}{E_\nu^2}
  \right) \ .
\end{equation}
Here $m_i$ refers to the isotope's atomic mass and $q_i^2=2m_iE_r$, where
$E_r^\text{max}\simeq 2E_\nu^2/m_i$, $E_\nu$ being the energy of the incoming neutrino.
 The overall energy-dependent
factor $\xi_V(q^2_i)$ encodes the CP violating physics and reads
\begin{equation}
  \label{eq:xiV}
  \xi_V(q^2_i)=g_V(q^2_i) - 
  \frac{h_V(q^2_i)(f_V - if_A)}{\sqrt{2}G_F(q^2_i-m_V^2)}\ ,
\end{equation}
with $g_V(q^2)$ the SM contribution weighted properly by the nuclear
form factors, namely
\begin{equation}
  \label{eq:SM-contribution}
  g_V(q^2_i)=N\left(2g_V^d+g_V^u\right)
    F_n(q^2)
    +
    Z\left(g_V^d+2g_V^u\right)F_p(q^2)\ ,
\end{equation}
with $g_V^u=1/2-4/3\,\sin^2\theta_W$ and
$g_V^d=-1/2+2/3\,\sin^2\theta_W$. For the weak mixing angle we use the
central value obtained using the $\overline{\text{MS}}$
renormalization scheme and evaluated at the $Z$ boson mass,
$\sin^2\theta_W=0.23122$ \cite{1674-1137-40-10-100001}.

Typical nuclear form factors parametrizations depend on two parameters
which are constrained via the corresponding nucleonic density
distribution root-mean-square (rms) radii. For a large range of
nuclides, proton rms radii have been precisely extracted from a
variety of experiments \cite{Angeli:2013epw}.  Consequently,
uncertainties on $F_p(q^2)$ are to a large degree negligible. In
contrast, neutron rms radii are poorly known and so uncertainties on
$F_n(q^2)$ can be large. These uncertainties have been recently
studied in \cite{AristizabalSierra:2019zmy} by assuming that
$r_\text{rms}^n\subset [r_\text{rms}^p,r_\text{rms}^p+0.3\,\text{fm}]$
(for heavy nuclei).  The lower bound is well justified in nuclides
with $N>Z$, while the upper one is limited by constraints from neutron
skin thickness \cite{Centelles:2008vu}. In our analysis we choose to
fix $r_\text{rms}^n=r_\text{rms}^p$ and use the same form factor
parametrization (Helm form factor \cite{Helm:1956zz}) for both,
neutrons and protons \cite{AristizabalSierra:2019zmy}. Doing so, the
$q^2$ dependence of the parameter in (\ref{eq:xiV}) comes entirely
from the denominator in the second term and that enables a
simplification of the multi-parameter problem. Note that we do not
consider form factor uncertainties in order to avoid mixing their
effects with the CP violating effects we want to highlight.

In general the analysis of CP violating effects is a nine parameter
problem: the vector boson mass, four moduli and four CP
phases. However, the problem can be reduced to three parameters by
rewriting the product of the nuclear and neutrino complex couplings
in the second term in (\ref{eq:xiV}) in terms of real and complex
components. A moduli
$|H_V|^2=\mathbb{R}\text{e}\,(H_V)^2+i\,\mathbb{I}\text{m}\,(H_V)^2$ a
phase $\tan\phi=\mathbb{I}\text{m}(H_V)/\mathbb{R}\text{e}(H_V)$ and
the vector boson mass. In terms of the fundamental couplings and CP
phases, they are given by
\begin{align}
  \label{eq:reparametrization-no-q-dependence}
  \mathbb{R}\text{e}\,(H_V)&=2(f_V+f_A)\sum_{q=u,d}\mathcal{A}_q\,h^q_V
                             \sin(\alpha/4)\,
                             \sin(\beta_+^q/4)\ ,
                             \nonumber\\
  \mathbb{I}\text{m}\,(H_V)&=2(f_A-f_V)\sum_{q=u,d}\mathcal{A}_q\,h^q_V
                             \sin(\alpha/4)\,
                             \sin(\beta_-^q/4)\ ,
\end{align}
with $\mathcal{A}_d=2A-Z$, $\mathcal{A}_u=Z+A$,
$\alpha=\pi+2(\phi_A-\phi_V)$ and
$\beta_\pm^q=\pi\pm 2(\phi_A+\phi_V+2\phi_{Vq})$. Proceeding in this
way the cross section then depends on $m_V$, $|H_V|$ and $\phi$
through the parameter $\xi_V$ in~(\ref{eq:xiV}), that is now
simplified to
\begin{equation}
  \label{eq:xi_V-simplified}
  \xi_V=g_V + \frac{|H_V|e^{i\phi}}{\sqrt{2}G_F\left(2m_iE_r+m_V^2\right)}\ .
\end{equation}
One can see that the cross section is invariant under $\phi\to -\phi$,
so the analysis can be done by considering $\phi\subset [0,\pi]$.
The phase reflection invariance of the cross section assures that the
results obtained for such interval hold as well for
$\phi\subset [-\pi,0]$. The boundaries of this interval define the two
CP conserving cases of our analysis. Since $g_V$ is always negative,
$\phi=0$ always produces destructive interference between the SM and
the light vector contribution. At the recoil spectrum level this
translates into a depletion of the SM prediction in a certain recoil
energy interval. In contrast, $\phi=\pi$ implies always constructive
interference, and so an enhancement of the recoil spectrum above the
SM expectation.

It becomes clear as well that the conclusions derived in terms of
$|H_V|$ and $\phi$ can then be mapped into the eight-dimensional
parameter space spanned by the set
$\{|f_{V,A}|,|h_V^q|,\phi_{V,A},\phi_{Vq}\}$.
\section{Event rates, constraints and parameter space analysis}
\label{sec:param-space-analysis}
To characterize CP violating effects we consider CE$\nu$NS produced by
fixed target experiments, in particular at COHERENT. Qualitatively,
the results derived here apply as well in the case of CE$\nu$NS
induced by reactor and solar ($^8$B) neutrinos. We start the analysis
by studying the effects in mono-target sodium
($^{23}$Na)\footnote{Throughout the paper we refer to this case as NaI
  detector. The high-energy $^{23}$Na recoils have a better
  signal-to-background ratio than $^{127}$I, and so CE$\nu$NS is
  measured in $^{23}$Na. Iodide is instead employed to measure
  $\nu_e-$induced charged current processes
  \cite{Akimov:2018vzs}. Thus, from the CE$\nu$NS point of view NaI is
  a mono-target experiment.}  and argon ($^{40}$Ar) detectors and then
consider the case of a multi-target germanium detector. For the latter
case one has to bear in mind that germanium has five stable isotopes
$^{70}$Ge, $^{72}$Ge, $^{73}$Ge, $^{74}$Ge and $^{76}$Ge with relative
abundances $20.4\%$, $27.3\%$, $7.76\%$, $36.7\%$ and $7.83\%$,
respectively.

In the multi-target case the contribution of the
$i^{\underline{\text{th}}}$ isotope to the energy recoil spectrum can
be written according to \cite{AristizabalSierra:2019zmy}
\begin{equation}
  \label{eq:recoil-spectrum-ith-isotope}
  \frac{dR_i}{dE_r}=\frac{m_\text{det}N_A}{\langle m\rangle}
  \int_{E_\nu^\text{min}}^{E_\nu^\text{max}}\Phi(E_\nu)X_i\frac{d\sigma_i}{dE_r}
  F_H^2(q^2_i)\,dE_\nu\ ,
\end{equation}
where $m_\text{det}$ is the detector mass in kg,
$\langle m\rangle=\sum_kX_km_k$ with $m_k$ the
$k^{\underline{\text{th}}}$ isotope molar mass measured in kg/mol,
$X_i$ is the isotope relative natural abundance,
$N_A=6.022\times 10^{23}\,\text{mol}^{-1}$, $\Phi(E_\nu)$ the
neutrino flux and $F_H^2(q^2_i)$ stands for the Helm form factor. 
The integration limits are $E_\nu^\text{max}=m_\mu/2$ (for a fixed-target experiment like COHERENT)
and $E_\nu^\text{min}=\sqrt{m_iE_r/2}$. The full recoil spectrum then
results from $dR/dE_r=\sum_i dR_i/dE_r$. Note that
(\ref{eq:recoil-spectrum-ith-isotope}) reduces to the single target
case when $X_i=1$ and $m=m_k=0.932 A_k\,\text{GeV}/\text{c}^2$. The
number of events in a particular detector is then calculated as
\begin{equation}
  \label{eq:events}
  N_\text{events}=\int_{E_r-\Delta E_r}^{E_r+\Delta E_r}\frac{dR}{dE_r}
  \mathcal{A}(E_r)dE_r\ ,
\end{equation}
with $\mathcal{A}(E_r)$ the acceptance function of the experiment. In
our analyses we take $\Delta E_r=1.5\,$keV.
\subsection{Constraints on light vector mediators}
\label{sec:contraints}
Before proceeding with our analysis it is worth reviewing the
constraints to which the light vector mediators we consider are
subject to. These constraints arise from beam dump and fixed target
experiments, $e^+e^-$ colliders and LHC, lepton precision experiments,
neutrino data as well as astrophysical observations
\cite{Bauer:2018onh}. From the collision of an electron or proton beam
on a fixed target, $V$ can be produced either through Bremsstrahlung
or meson production and subsequent decay, $\pi^0\to \gamma + V$. The
interactions in (\ref{eq:BSM-Lag}) do not involve charged leptons, hence
in the light mediator scenario here considered the coupling of
$V$ to electrons is loop suppressed. Limits from electron beam dump
and fixed target experiments can be therefore safely ignored. Limits
from proton beams are seemingly more relevant since the production of $V$
is possible by Bremsstrahlung---through the vertex
$\bar p\gamma_\mu p V^\mu$---or by meson decay. However, since these
searches are based on $V$ decay modes involving charged leptons, again
the constraints are weaken by loop suppression factors.
\begin{figure*}
  \centering
  \includegraphics[scale=0.45]{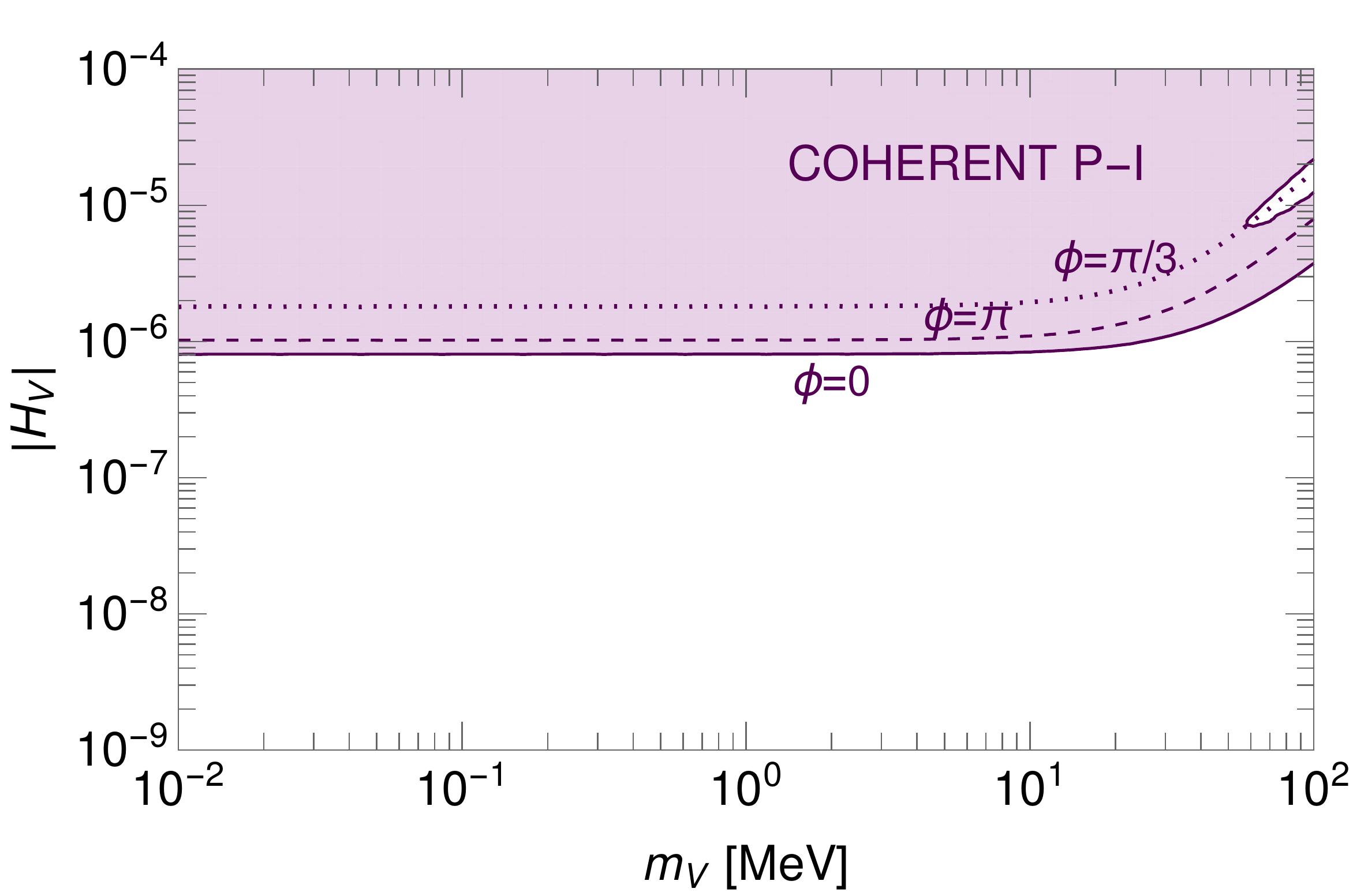}
  \caption{Bounds on vector light mediators derived from COHERENT P-I
    data. The bounds include the real cases $\phi=0$ and $\phi=\pi$ as
    well as $\phi=\pi/3$, value for which the limit is found to be the
    less stringent.}
  \label{fig:coherent-PI}
\end{figure*}

The potential limits from $e^+e^-$ collider searches (e.g. KLOE, BaBar
or Belle-II \cite{Anastasi:2016ktq,Lees:2014xha,Inguglia:2016acz}),
from muon and tau rare decays (SINDRUM and CLEO
\cite{Bertl:1985mw,Alam:1995mt}) and from LHC searches (LHCb, ATLAS
and CMS \cite{Aaij:2017rft,Curtin:2014cca}) are feeble due to the same
argument, couplings of $V$ to charged leptons are loop suppressed. As to
the limits from neutrino scattering experiments, Borexino, neutrino
trident production and TEXONO
\cite{Harnik:2012ni,Altmannshofer:2014pba,Bilmis:2015lja} involve
couplings to charged leptons and so are weak too. Thus, from
laboratory experiments the only relevant limit arises from COHERENT
CsI phase \cite{Akimov:2017ade}, which have been studied in detail in
ref. \cite{Liao:2017uzy} under the assumption of real parameters. We
thus update those limits by considering $\phi\neq 0$.  To do so we
follow the same strategy adopted in
ref. \cite{AristizabalSierra:2018eqm}. First of all, we define the
following spectral $\chi^2$ function
\begin{align}
  \label{eq:chiSq}
  \chi^2&=\sum_{i=1}^{16}\left(\frac{N^\text{exp}_i-(1+\alpha)N^\text{BSM}_i
          -(1+\beta)B_i^\text{on}}{\sigma_i}\right)^2
          \nonumber\\
        &+ \left(\frac{\alpha}{\sigma_\alpha}\right)^2
          + \left(\frac{\beta}{\sigma_\beta}\right)^2\ ,
\end{align}
where the binning runs over number of photoelectrons $n_\text{PE}$
($\Delta n_\text{PE}=2$ and $n_\text{PE}=1.17(E_r/\text{keV})$),
$\alpha$ and $\beta$ are nuisance parameters, $\sigma_i$ are
experimental statistical uncertainties and $\sigma_\alpha=0.28$ and
$\sigma_\beta=0.25$ quantify standard deviations in signal and
background respectively. For the calculation of $N^\text{BSM}_i$ we
employ eqs. (\ref{eq:recoil-spectrum-ith-isotope}) and
(\ref{eq:events}) adapted to include the Cs and I contributions,
i.e. $m_\text{det}=14.6\,$kg, $\langle m\rangle\to m_\text{CsI}$
($m_\text{CsI}$ the CsI molar mass) and
$X_i\to A_i/(A_\text{Cs}+A_\text{I})$. For neutrino fluxes we use the
following spectral functions
\begin{align}
  \label{spectral-functions}
  \mathcal{F}_{\nu_\mu}(E_\nu)&=\frac{2m_\pi}{m_\pi^2-m_\mu^2}
                                \delta\left(1-\frac{2E_\nu m_\pi}{m_\pi^2-m_\mu^2}\right)\ ,
                                \nonumber\\
  \mathcal{F}_{\nu_e}(E_\nu)&=\frac{192}{m_\mu}\left(\frac{E_\nu}{m_\mu}\right)^2
                              \left(\frac{1}{2}-\frac{E_\nu}{m_\mu}\right)\ ,
                              \nonumber\\
  \mathcal{F}_{\bar\nu_\mu}(E_\nu)&=\frac{64}{m_\mu}\left(\frac{E_\nu}{m_\mu}\right)^2
                                    \left(\frac{3}{4}-\frac{E_\nu}{m_\mu}\right)\ ,
\end{align}
normalized according to $\mathcal{N}=r\times n_\text{POT}/4/\pi/L^2$,
with $r=0.08$, $n_\text{POT}=1.76\times 10^{23}$ and $L=19.3\,$m. The
result is displayed in fig.~\ref{fig:coherent-PI} where it can be seen
that the inclusion of CP phases relaxes the bound.  We found that the
less stringent limit is obtained for $\phi=\pi/3$, which is about a
factor 2.5 larger than the bound obtained at $\phi=0$.

The last limits which apply in our case are of astrophysical
origin. Particularly important are horizontal branch stars which have
a burning helium core with
$T\simeq 10^8\,\text{K}\simeq 10^{-2}\,\text{MeV}$. In such an
environment vector bosons with masses of up to $10^{-1}\,$MeV (from
the tail of the thermal distribution) can be produced through Compton
scattering processes $\gamma+^4\text{He}\to V+^4\text{He}$ which lead
to energy loss. Consistency with the observed number ratio of
horizontal branch stars in globular clusters leads to a constraint on
the vector-nucleon couplings $h_V^{p,n}\lesssim 4\times 10^{-11}$
\cite{Grifols:1986fc,Grifols:1988fv}. Assuming $h_V^p=h_V^n$ this
bound can be translated into
$|H_V|= \sqrt{2}A h_V^n\lesssim 6\,\times 10^{-11}\,A$. Relevant as
well are the bounds derived from supernova, which exclude regions in
parameter space for light vector boson masses up to
$\sim 100\,$MeV\footnote{Supernova temperatures are order
  $T\simeq 30\,$MeV, and so in that environment states with masses of
  up to $\sim 100\,$MeV can be produced if one consider the tail of
  the thermal distribution \cite{Chang:2018rso}.}. Neutrinos are
trapped in the supernova core, so they can only escape by
diffusion. Consistency with observations implies
$t_\text{diff}\sim 10\,$s, therefore limits can be derived by
requiring that the new interaction does not sizably disrupt
$t_\text{diff}$. Further limits can be derived from energy-loss
arguments if the new interactions open new channels for neutrino
emission, which is the case in the scenario we are considering
through $V\to \bar\nu\nu$ (a process that resemble the plasma process
$\gamma\to \bar\nu\nu$). All these limits have been recently reviewed
for dark photons in \cite{Chang:2016ntp} and span a region of
parameter space that covers several orders of magnitude in both
$|H_V|$ and $m_V$.

\begin{figure*}
  \centering
  \includegraphics[scale=0.445]{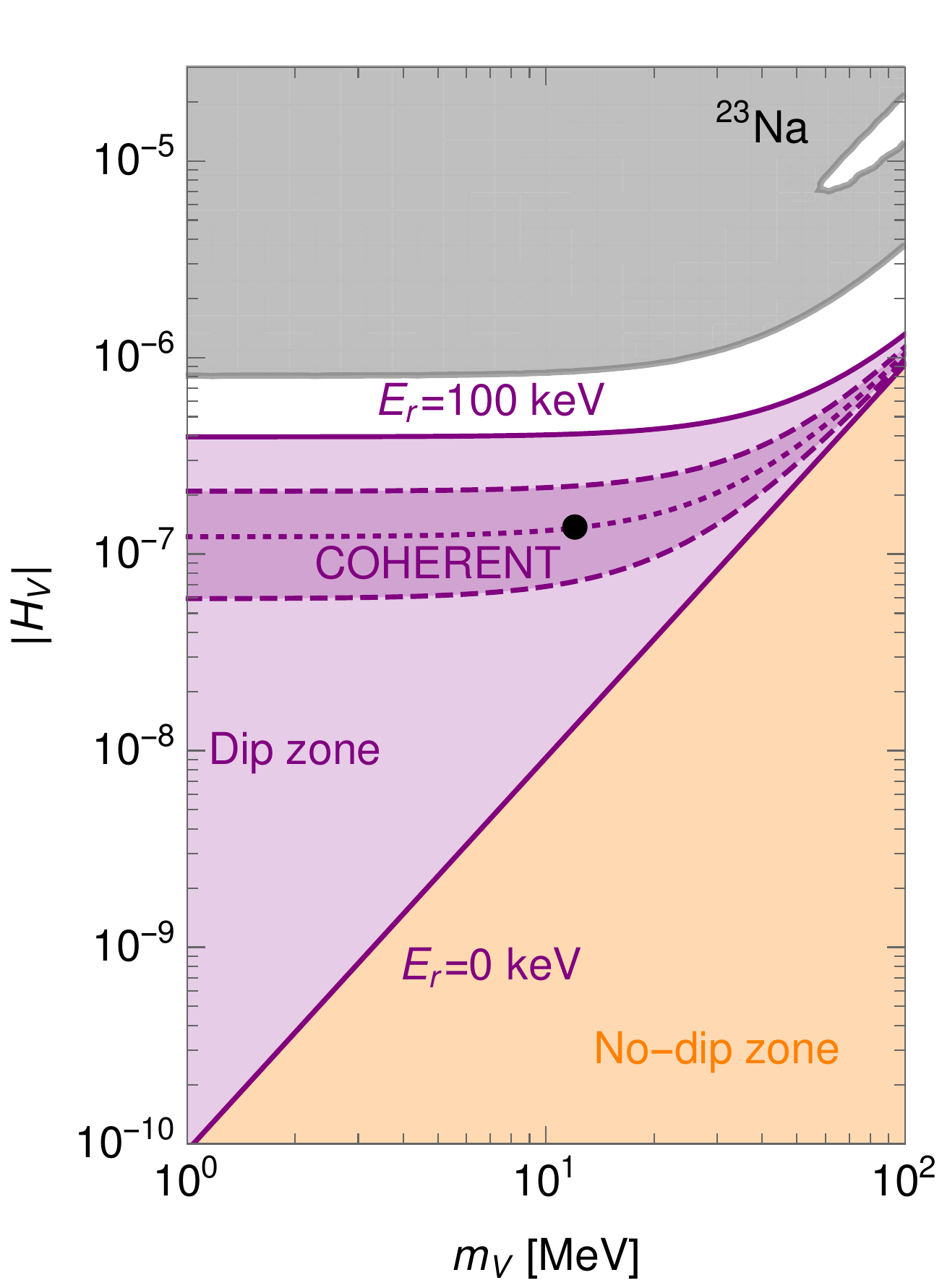}
  \includegraphics[scale=0.445]{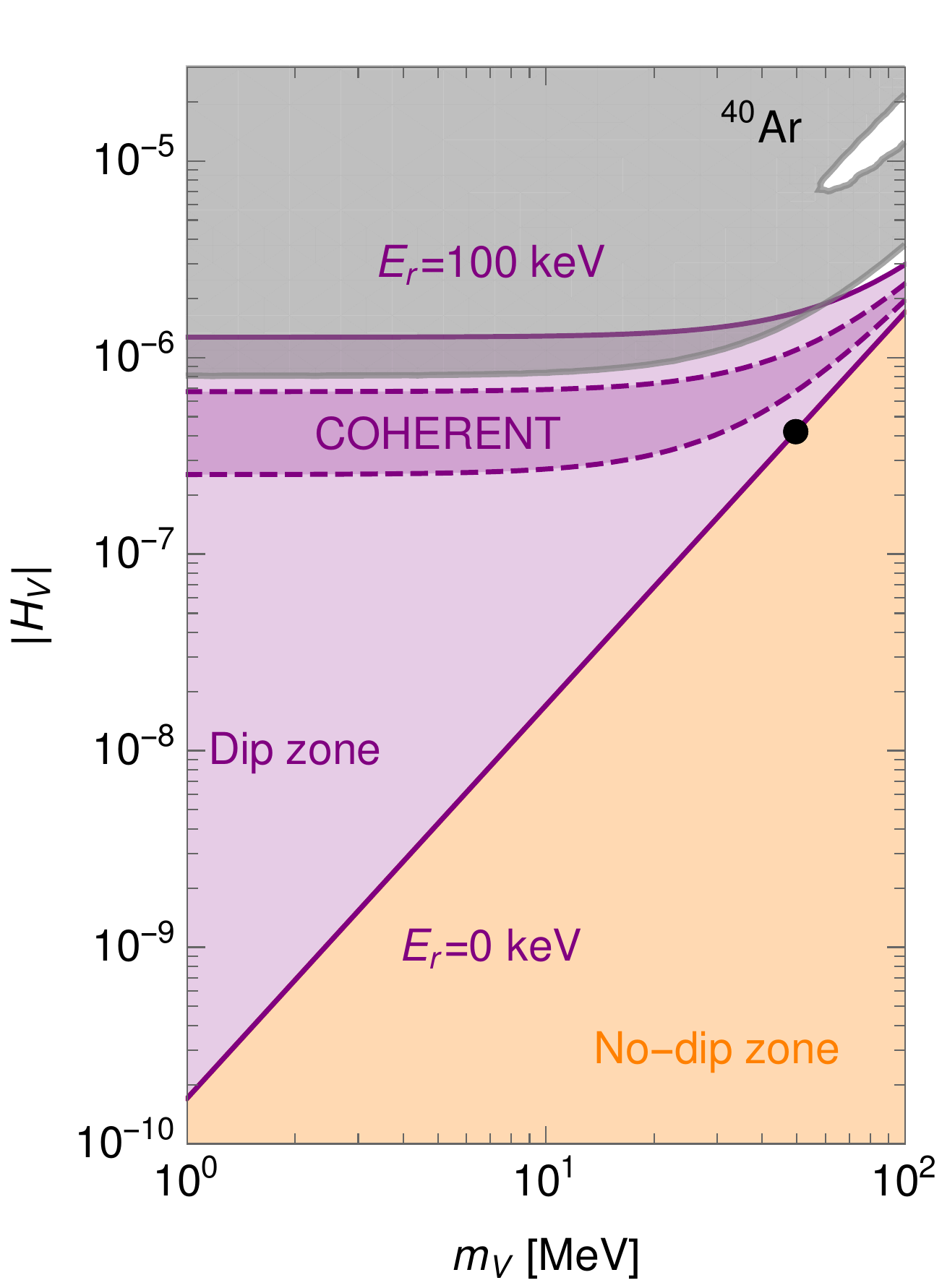}
  \includegraphics[scale=0.445]{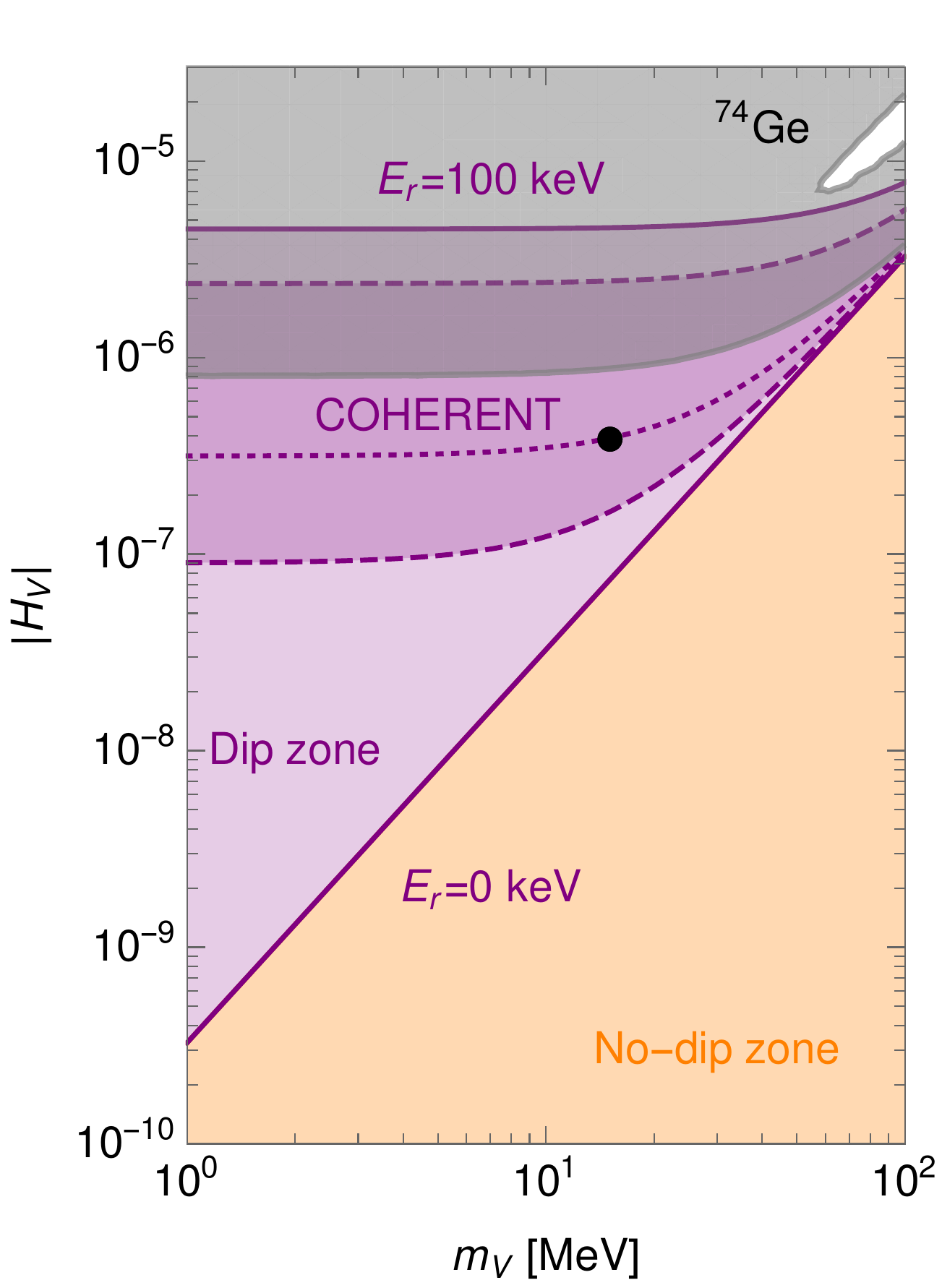}
  \caption{Parameter space regions determined by the condition that
    the spectrum exhibits or not a dip at a given recoil energy, in
    $^{23}$Na (\textbf{left graph}), $^{40}$Ar (\textbf{middle graph})
    and $^{74}$Ge (\textbf{right graph}). For parameters in the upper
    triangle (purple region) a dip in the event rate spectrum is
    always found, while this is not the case for parameters in the
    lower triangle (orange region). The plain and dashed/dotted curves are isocontours referring to increasing values of recoil energy, from $E_r = 0$ (plain diagonal purple line) to $E_r = 100$ keV (top plain purple line). The shaded regions labeled with
    COHERENT indicate the recoil energy regions of interest for the
    sodium, argon and germanium COHERENT detectors (see
    sec. \ref{sec:param-space} for further details). The grey shaded area indicates the region already 
    excluded by COHERENT P-I, assuming the real case $\phi = 0$ (see Fig.~\ref{fig:coherent-PI}). }
  \label{fig:triangles}
\end{figure*}
There are various considerations that have to be taken into account
regarding these bounds. First of all, uncertainties on the behavior of
core-collapse supernovae are still substantial
\cite{Muller:2016izw}. As a result, limits from supernovae should be
understood as order-of-magnitude estimations. The bounds from stellar
cooling arguments discussed above neglect plasma mixing effects, which
are relevant whenever the vector has an effective in-medium mixing
with the photon. Taking into account these effects, the
production rate of the new vectors in the stellar environment is affected,
resulting in rather different bounds \cite{Hardy:2016kme}. Additional
environmental effects can alter the bounds from stellar cooling as
well as from supernova. This is the case when the vector couples to a
scalar which condensates inside macroscopic objects, and screens the
charge which $V$ couples to\cite{Nelson:2007yq,Nelson:2008tn}. The
vector mass in this scenario is proportional to the medium mass
density $\rho$, and so in stellar and supernova environments
(high-density environments) its production is no longer possible. In
summary, astrophysical constraints should be considered with care as
they largely depend on the assumptions used. Thus, for concreteness
and because this is the window where new CP violating effects are more
pronounced, we focus our analysis in the region
$m_V\subset [1,100]\,$MeV.
\subsection{Parameter space slicing}
\label{sec:param-space}
For CP conserving parameters a full cancellation of the SM
contribution, at a given recoil energy, becomes possible in the case
$\phi=0$. In contrast, CP violating parameters do not allow such a
possibility. For $N_\text{events}$ such a cancellation leads to a dip
at the recoil energy at which the cancellation takes place. Thus, such
a feature in the spectrum will favor CP conserving new physics. Taking
this into account, we then split the $m_V-|H_V|$ plane in two
``slices'': One for which the recoil spectrum will always exhibit a
dip, and a second one for which this is never the case, regardless of
$\phi$. The boundary of such regions is clearly determined by the
condition that the parameter in eq. (\ref{eq:xi_V-simplified})
vanishes, which translates into a relation between $|H_V|$ and $m_V$
for a fixed recoil energy, namely
\begin{equation}
  \label{eq:Hv-mV}
  |H_V|=-\sqrt{2}g_VG_F\left(2m_iE_r + m_V^2\right)\ .
\end{equation}
In a mono-target experiment the cancellation is exact at a given
energy, but in a multi-target detector this is clearly not the
case. However, as we will later show in
sec. \ref{sec:sodium-germanium} the cancellation is still good enough
so to be used to distinguish the CP conserving case from the CP
violating one. One can see as well that the position of the dips
implied by eq.~(\ref{eq:Hv-mV}) depends on the type of isotope
considered, so different nuclides span different portions of parameter
space. This can be seen in fig. \ref{fig:triangles} in which the
parameter space regions $m_V-|H_V|$ are displayed for $^{23}$Na,
$^{40}$Ar and $^{74}$Ge.

\begin{table*}
  \renewcommand{\arraystretch}{1.4}
  \setlength{\tabcolsep}{7pt}
  \centering
  \begin{tabular}{|c||c|c|c|}\hline
    \textbf{Detector}&\textbf{Detector mass [kg]}& \textbf{Distance from source [m]}& \textbf{Threshold [keV]}
    \\\hline
    Sodium           & $2000$                                 &               $28$         &   $15$\\\hline
    Liquid Argon     & $1000$                                 &               $29$         &   $20$\\\hline
    Germanium        & $15$                                   &               $22$         &   $2$\\\hline
  \end{tabular}
  \caption{Main detector parameters employed in our analysis, taken
    from \cite{scholberg}. For the number of protons on target $n_\text{POT}$ we have
    extrapolated the value of the CsI phase to one year in all cases, 
    $n_\text{POT}=(365/308.1) 1.76\times 10^{23}=2.1\times 10^{23}$. Acceptances $\mathcal{A}(E_r)$
    are taken as Heaviside functions at the energy thresholds specified in the fourth column.}
  \label{tab:parameters-detectors}
\end{table*}
The regions labeled with COHERENT refer to the energy regions of
interest in each case. In all three cases the upper energy isocontour
is fixed as $E_r=50\,$keV (determined by the $\nu_e$ flux kinematic
endpoint), and the lower isocontour according to the projected
detector recoil energy thresholds. For the NaI detector we assume
$E_r^\text{th}=15\,$keV, for the LAr $E_r^\text{th}=20\,$keV and for
germanium $E_r^\text{th}=2\,$keV. The lower isocontour at $E_r=0\,$keV
defines the boundary of the regions with distinctive and not
overlapping CP violating features: dips and degeneracies. The upper
isocontour at $E_r=100\,$keV is fixed by the condition of keeping the
elastic neutrino-nucleus scattering coherent.  Apart from these
particular energy isocontours, any other one within the dip zone determines
the position of the dip. This means that if future data will show a dip
in the event spectrum, and one interprets such a dip in terms of a light
vector mediator scenario, its energy location will provide valuable information
about the new physics parameters.

\begin{figure*}
  \centering
  \includegraphics[scale=0.373]{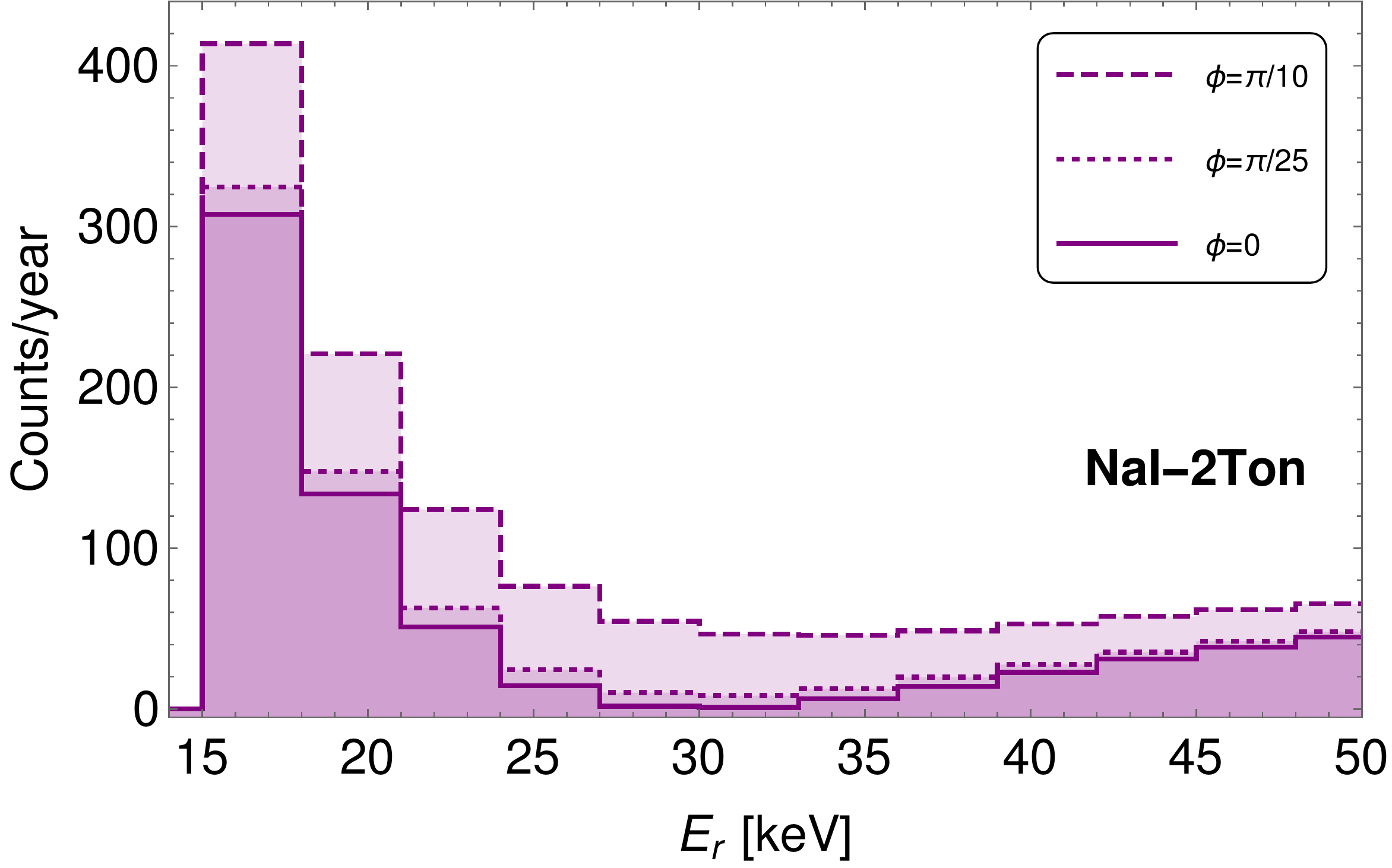}
  \hfill
  \includegraphics[scale=0.373]{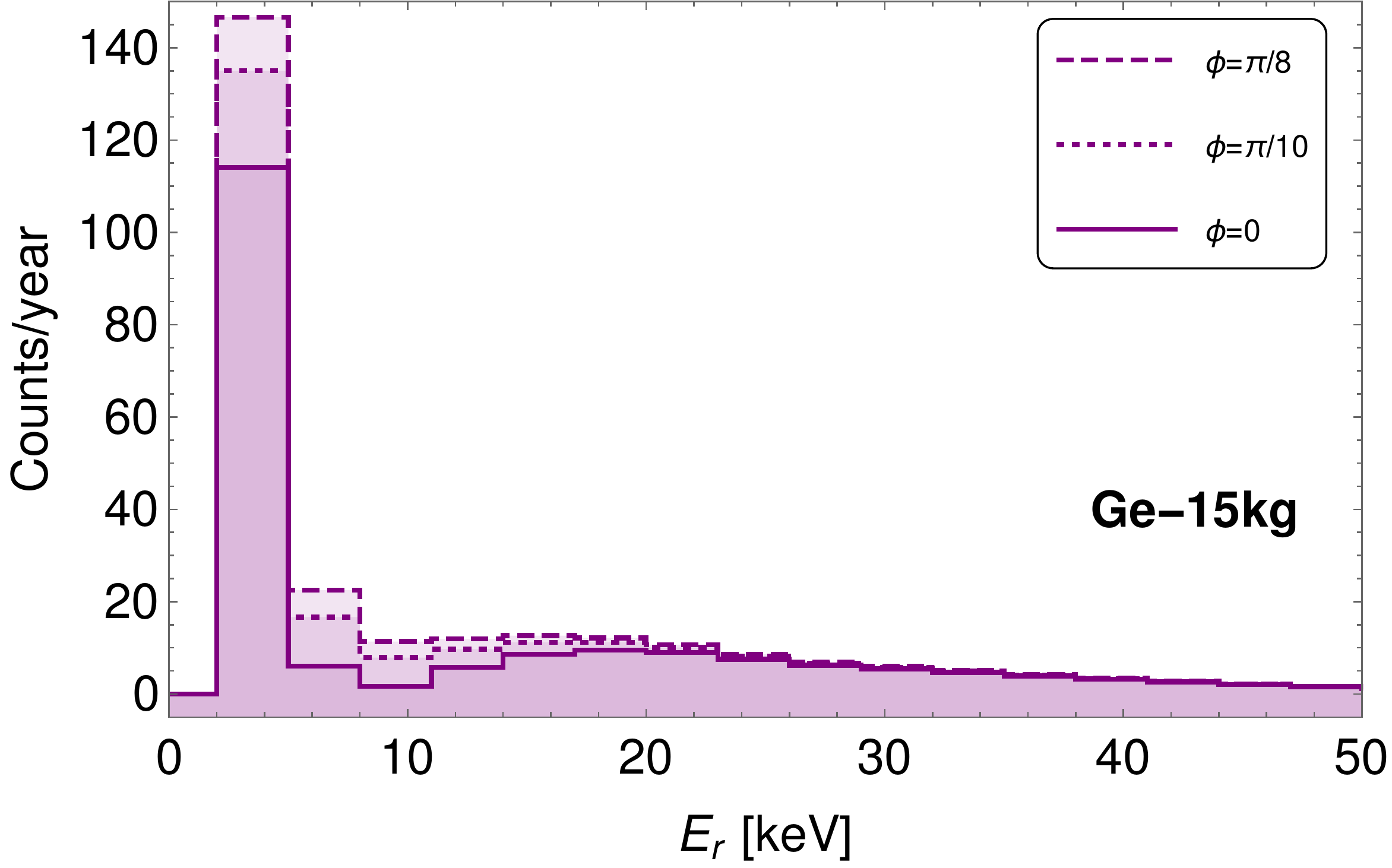}
  \includegraphics[scale=0.373]{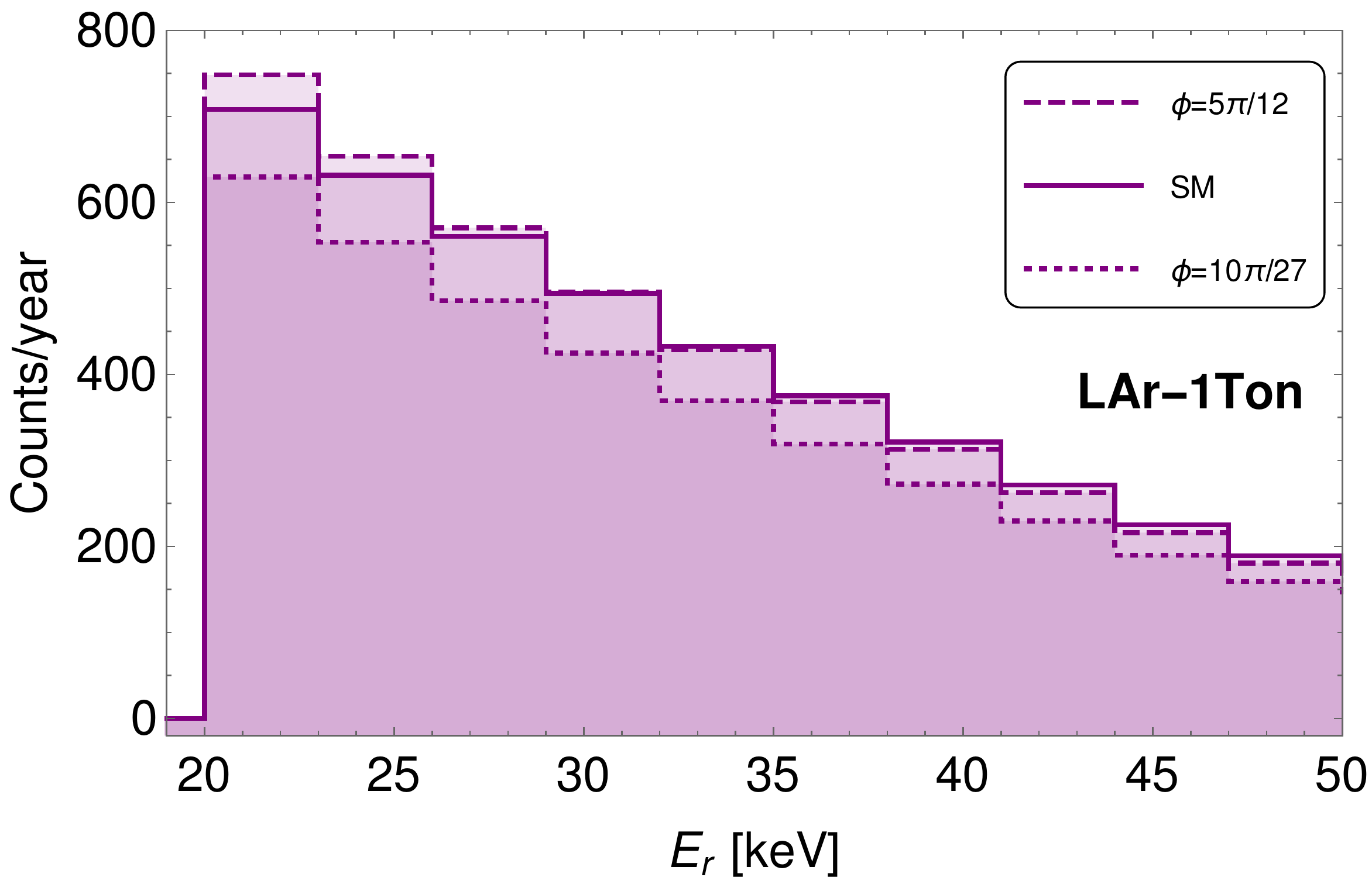}
  \hfill
  \includegraphics[scale=0.373]{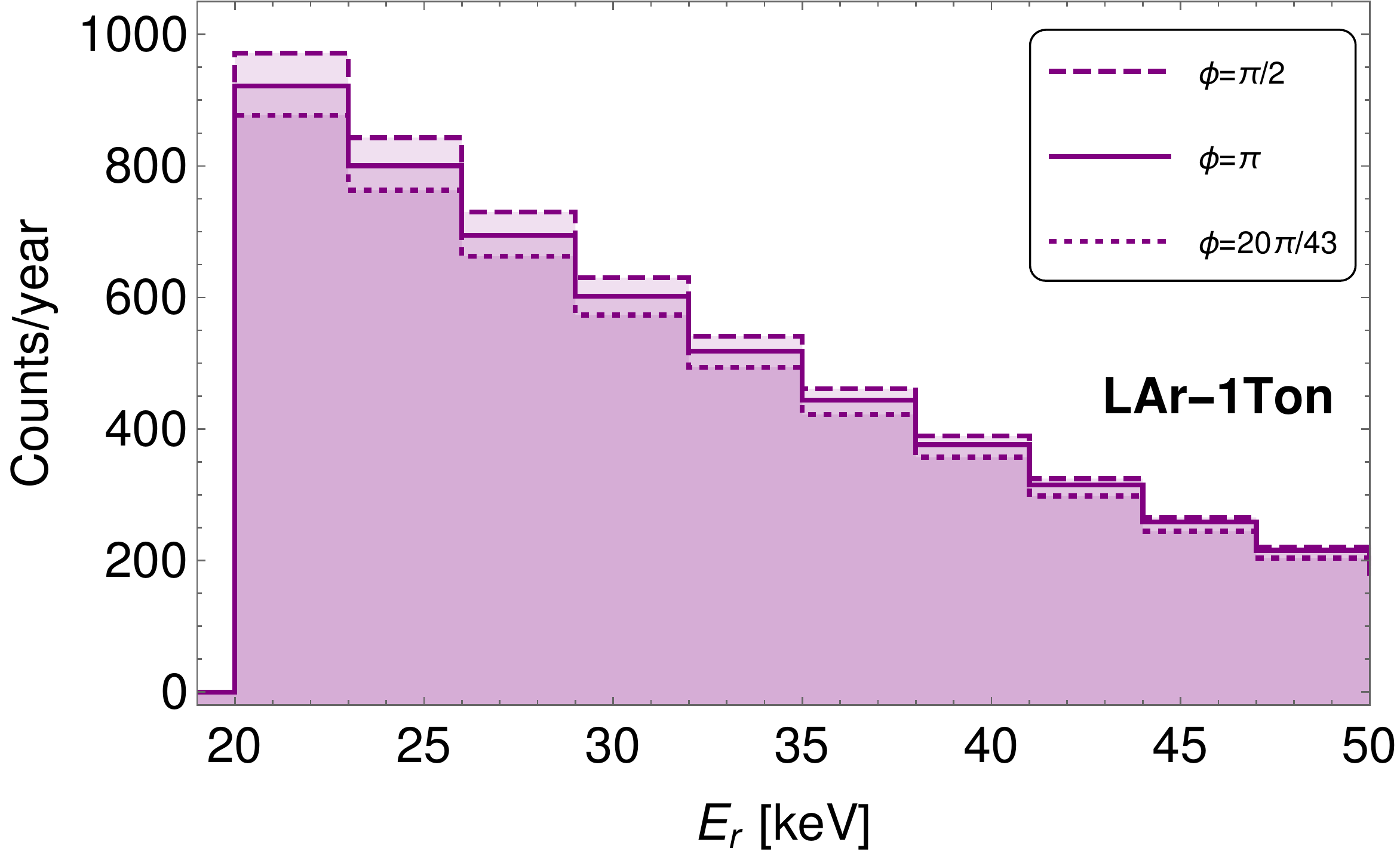}
  \caption{\textbf{Upper left}: Event rate spectrum calculated in the
    NaI detector assuming one-year exposure and
    $m_\text{det}=2\,$ton. The dip is obtained for $\phi=0$, the
    presence of CP violation lifts the dip in such a way that the
    larger $\phi$ the less pronounced the dip. Observation of these
    features can then be used to constrain the amount of CP violation involved by
    the new physics. \textbf{Upper right}: Same as in the NaI
    detector, but calculated for the germanium detector assuming full
    cancellation in $^{74}$Ge and $m_\text{det}=15\,$kg. In this case
    the similarity of the five isotopes leads to dips for all of them
    that spread within a recoil energy interval of $2\,$keV. The set
    of dips within that narrow window leads to an overall dip as shown
    in the graph. \textbf{Lower left}: Degeneracy between the SM
    prediction and the SM+vector with $\phi\neq 0$ in the LAr detector
    with $m_\text{det}=1\,$ton. \textbf{Lower right}: Degeneracy
    between a spectrum generated with real parameters and spectra
    generated with $\phi\neq 0$ in the LAr detector as well. These two
    cases, labeled SM degeneracy and real-vs-complex degeneracy, show
    that the interpretation of a CE$\nu$NS signal should be done
    including CP violating effects.}
  \label{fig:dips-and-degeneracy}
\end{figure*}

To emphasize this observation we consider the $^{23}$Na mono-target
detector as well as the germanium multi-target detector. In the first
case, we consider the parameter space point
$\{m_V,|H_V|\}=\{12\,\text{MeV},1.32\times 10^{-7}\}$ as indicated in
the left panel of fig. \ref{fig:triangles} with a black point. That point is
located along the $E_r=31\,$keV dotted isocontour, so with $\phi=0$ a
dip in that position is found as shown in the upper left graph in
fig. \ref{fig:dips-and-degeneracy} (detector parameters used for this
calculation can be seen in tab. \ref{tab:parameters-detectors}). Data
from that detector will identify its exact location up to bin size
(energy resolution). Assuming $\Delta E_r=1.5\,$keV, such a spectrum will
allow to determine $|H_V|$ with a $4\%$ accuracy within the range
$[1.22\times 10^{-7},1.04\times 10^{-6}]$ obtained at $m_V=1\,$MeV and
$m_V=100\,$MeV, respectively.

As the upper left panel in fig. \ref{fig:dips-and-degeneracy} shows,
the presence of CP violating phases produces departures from the dip
and so---in principle---one can relate the amount of CP violation to
the dip depth. In a mono-target detector this behavior is rather clear
given that the dip is related with a cancellation in a single
isotope. In a multi-target detector such as for germanium this is not
entirely clear. So let us discuss this in more detail. The event rate
spectrum is obtained from five different contributions, according to
eq. (\ref{eq:recoil-spectrum-ith-isotope}). Cancellation at a certain
recoil energy for a specific isotope requires a precise value of $H_V$
determined by the isotope mass and mass number, and so one expects the
remaining contributions not to cancel at that energy.

To investigate what happens in this case, we take the
parameter space point
$\{m_V,|H_V|\}=\{15\,\text{MeV},4.17\times 10^{-7}\}$, located along
the $E_r=7\,$keV isocontour for $^{74}$Ge, as indicated in the right
graph in fig. \ref{fig:triangles} with the black point. For that
point, the quantity
$\widehat \sigma_i=X_i\,(d\sigma_i/dE_r)\,F_H^2(q_i)$ exactly cancels
for $^{74}$Ge and $E_\nu=50\,$MeV (any other value allowed by the
kinematic criterion $E_\nu>\sqrt{m_iE_r/2}$ will lead to the same
conclusion). For the remaining isotopes, instead, the following values
are found
\begin{alignat}{2}
  \label{eq:sigmahatforgermaniun}
  \widehat\sigma_{70}&=1.5\times 10^{-39}\;\frac{\text{cm}^2}{\text{MeV}}\ ,&
  \quad\widehat\sigma_{72}=5.1\times 10^{-40}\;\frac{\text{cm}^2}{\text{MeV}}\ ,
  \nonumber\\
  \widehat\sigma_{73}&=3.6\times 10^{-41}\;\frac{\text{cm}^2}{\text{MeV}}\ ,&
  \quad\widehat\sigma_{76}=1.4\times 10^{-40}\;\frac{\text{cm}^2}{\text{MeV}}\ ,
\end{alignat}
which certainly are rather sizable. The key observation here is that
for the same parameter space point all five isotopes generate a dip
within a recoil energy interval of $2\,$keV. More precisely, at
$E_r=8.4\,$keV, $E_r=7.6\,$keV, $E_r=7.3\,$keV, $E_r=6.4\,$keV for
$^{70}$Ge, $^{72}$Ge,$^{73}$Ge, $^{76}$Ge respectively. Thus, given
the spread of those dips, the event rate spectrum does involve a
rather pronounced depletion that looks like the dip found in a
mono-target detector.

Note that the reason behind the appearance of multiple dips from
different germanium isotopes has to do with their similarity. The
value of $|H_V|$ for a fixed vector boson mass is entirely determined
by $m_i$ and $A_i$ through eq. (\ref{eq:Hv-mV}). Once the value of
$|H_V|$ is fixed using the mass and mass number of a particular
isotope (in this particular case $^{74}$Ge), eq. (\ref{eq:Hv-mV})
fixes as well the points at which the remaining dips will appear. The
different recoil energy positions differ only by the relative values
of $g_V^i$ and $m_i$ compared to those of the isotope that has been
used to fix $|H_V|$. For $^{70}$Ge these differences are order $10\%$
and $5\%$, while for $^{76}$Ge they are $5\%$ and $2\%$. Since the
differences for $^{70}$Ge are the largest, for this isotope
one finds the largest shift from $E_r=7\,$keV. Moreover, since the
differences in all cases are small, the spread of the dips is small as
well. This conclusion is therefore independent of the parameter space
point chosen: \textit{There exists as well a dip zone in a multi-target detector 
(in this case, Germanium based), for which given a point in it the event rate
  spectrum will always exhibit a dip.}

This behavior can be seen in the upper right graph in
fig. \ref{fig:dips-and-degeneracy}. The overall dip is a result of the
five contributions and of their dips spreading over a small recoil
energy window around $\sim7\,$keV. One can see as well that the
presence of CP violating phases has the same effect that in a
mono-target detector. As soon as they are switched on, departures from
the dip are seen, and the behavior is such that large $\phi$ tends to
soften the dip. At this point it is therefore clear that in both,
mono- and multi-target detectors one could expect a dip which provides
information about whether the new vector boson physics involves CP
violating phases and---eventually---allows to extract information about
its size. We have stressed that in a mono-target detector the exact
position of the dip allows for the reconstruction of the coupling $|H_V|$,
within an interval. The small spread of the dips for the different
germanium isotopes allows the same reconstruction procedure in the
multi-target case. An observation of a dip in the event rate spectrum
will fix the value of $|H_V|$ within an energy recoil isocontour up to
the recoil energy resolution, in the NaI, Ge and LAr detectors.

We now turn to the discussion of the ``no-dip zone'' regions in the
graphs in fig.~\ref{fig:triangles}. For that purpose we use the LAr
detector (middle graph and detector parameters according to
tab. \ref{tab:parameters-detectors}). As we have already mentioned,
the observation of a dip places the possible parameters responsible
for a signal within the upper triangles in the graphs in
fig. \ref{fig:triangles}. The question is then what are the
consequences of CP violating phases if the parameters $\{m_V,|H_V|\}$
are located in the lower triangular region. With $\phi=0$ only the SM
spectrum or a non-SM spectrum with real parameters can be
generated. In the first case one would like to know whether a SM-like
signal suffices to discard CP violation. In the second case, instead,
what can be said about $\phi$ from such a signal.

We generate the SM signal by fixing $|H_V|=0$ and then generate a set
of signals using the parameter space point
$\{m_V,|H_V|\}=\{50\,\text{MeV},4.25\times 10^{-7}\}$ for different values of
$\phi$, as shown in the lower left graph in
fig.~\ref{fig:dips-and-degeneracy}. The value for $|H_V|$ is obtained
by fixing $m_V=50\,$MeV in eq. (\ref{eq:xi_V-simplified}) at
$E_r=0\,$keV. In general, for a point in either the boundary of the
two regions or in the lower triangle the resulting spectra are rather
different from the SM prediction. However, we find that for suitable
values of $\phi$ one can always find SM+vector spectra that degenerate
to a large degree with that of the SM, as illustrated in the graph for
$\phi=5\pi/12$ and $\phi=10\pi/27$. Thus, we conclude that the
observation of a SM-like signal cannot be used to rule out CP
violating interactions.

We then fix a spectrum generated with real parameters with the point
$\{m_V,|H_V|\}=\{16\,\text{MeV},4.45\times 10^{-8}\}$ and $\phi=\pi$. As in the
previous case we try to find spectra that degenerate with this
one. For the point $\{50\,\text{MeV},4.25\times 10^{-7}\}$ (used in
the case of SM degeneracy as well), we find that $\phi=\pi/2$ and
$\phi=20\pi/43$ generate spectra that follow rather closely the ``real
spectrum''. In summary, therefore, in the no-dip zone we find that the
presence of CP violation leads to degeneracies that call for the
inclusion of CP violating effects if CE$\nu$NS data is to be
interpreted in terms of light vector mediators.
\begin{figure*}
  \centering
  \includegraphics[scale=0.373]{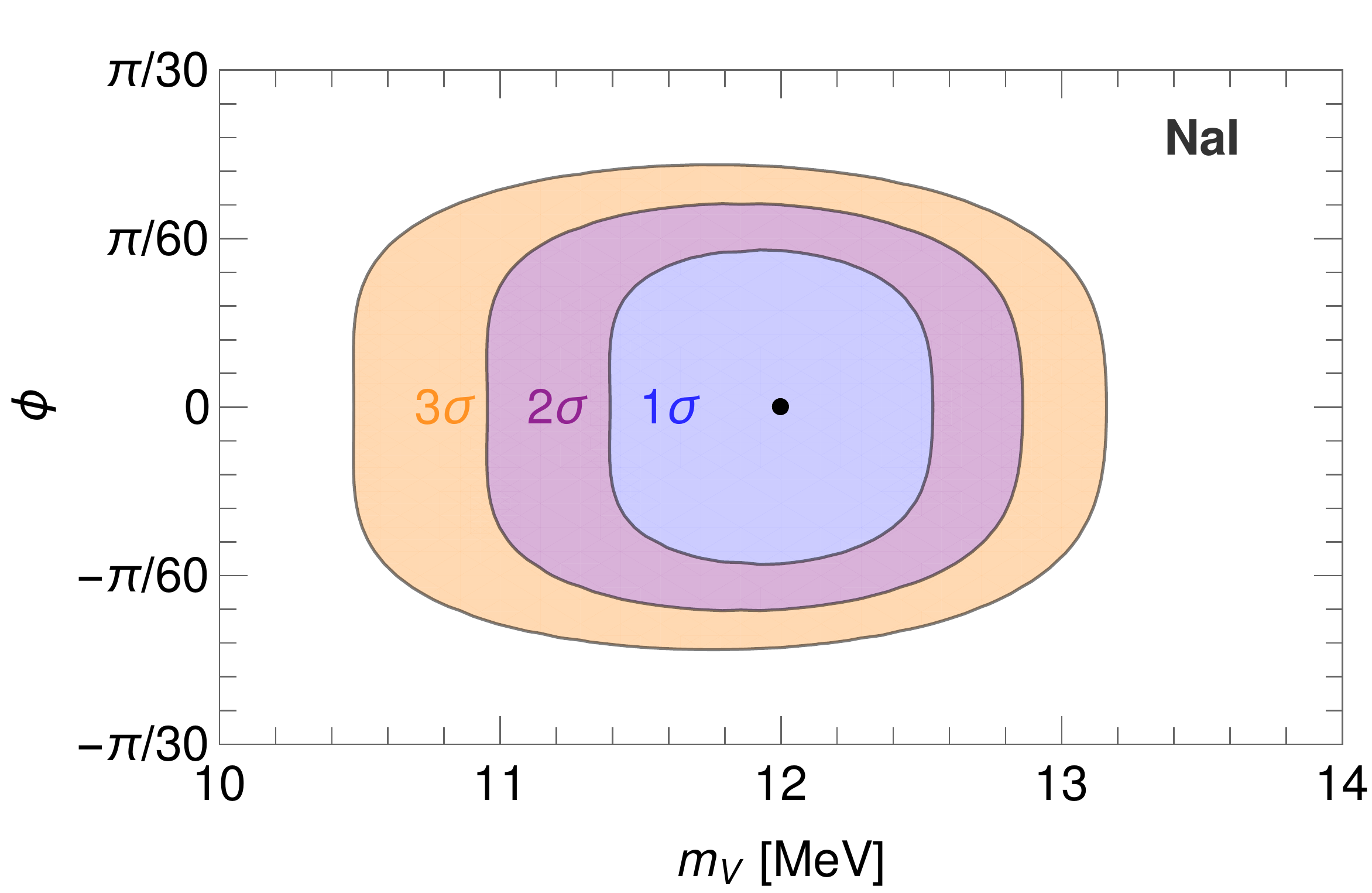}
  \hfill
  \includegraphics[scale=0.373]{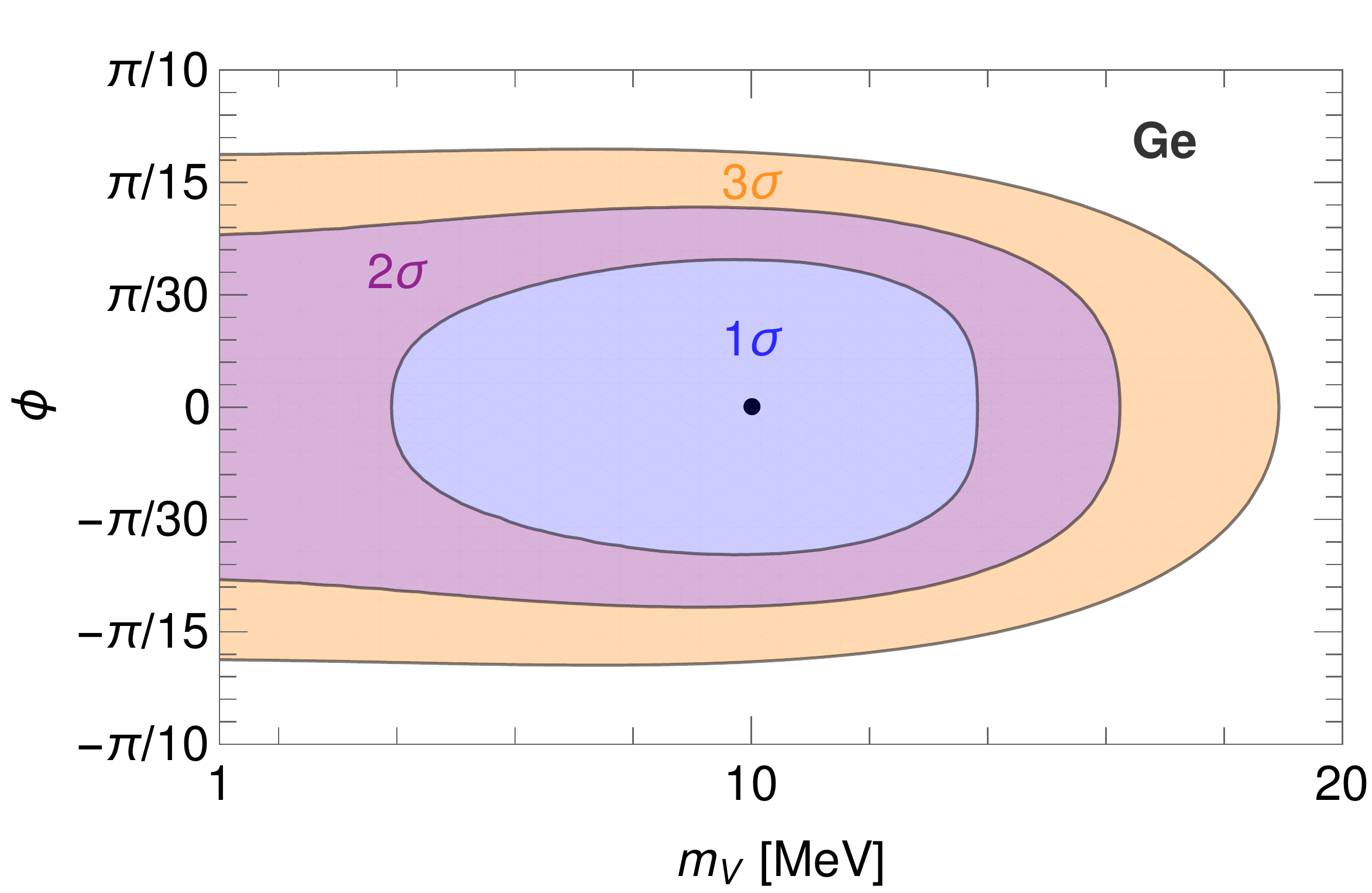}
  \caption{\textbf{Left graph}: Results of the chi-square analysis for
    the NaI detector. The three regions correspond to the 1$\sigma$,
    2$\sigma$ and 3$\sigma$ CL isocontours in the $m_V-\phi$
    plane. The contours show the degree at which a CE$\nu$NS signal
    involving a dip will allow to constrain the amount of CP
    violation. \textbf{Right graph}: Same as in the left graph but for
    the multi-target germanium detector. In this case the constraints on $\phi$,
    although still rather competitive, are less pronounced that in the
    NaI detector due to differences in the detector volume size. The
    black points indicate the best fit point value.}
  \label{fig:chisq-analysis-NaI-Ge}
\end{figure*}
\section{Determining the size of CP violating effects}
\label{sec:CPV-effects}
We have shown that the inclusion of CP violation has three main
effects: (i) suppression of eventual dips in the event rate spectrum,
(ii) degeneracy between the SM prediction and the light vector
mediator signal (SM degeneracy), (iii) degeneracy between spectra
generated with real parameters and spectra including CP violating
phases (real-vs-complex degeneracy). In what follows we study these
three cases in more detail.  We do so by taking four data sets that we
treat as pseudo-experiments. With them we then perform a $\chi^2$
analysis to show how much $\phi$ can be constrained with experimental
data. We assume a Poissonian distribution for the binned statistical
uncertainty, and so we do not include any steady-state nor beam-on
backgrounds.
\subsection{The case of sodium and germanium detectors}
\label{sec:sodium-germanium}
To show the degree at which the presence of a dip can constrain the
values of $\phi$, we do a counting experiment and perform a $\chi^2$
analysis. For that we employ eq.~(\ref{eq:chiSq}) considering only the
signal nuisance parameter and experimental signal uncertainty
$\sigma_\alpha$, which we keep as in the COHERENT CsI phase. In both
cases we use the neutrino fluxes from eq. (\ref{spectral-functions})
and we fix the remaining parameters according to
tab. \ref{tab:parameters-detectors}. For the NaI detector we use
$H(E_r/\text{keV}-15)$, while for the germanium detector
$H(E_r/\text{keV}-2)$. The binning is done in such a way that the
first data point is centered at $E_r^\text{th}/\text{keV}+1.5$.

For the NaI analysis, the data points used for $N_\text{exp}$ are
obtained by fixing $\phi=0$ and the parameter space point shown in the
left graph of fig. \ref{fig:triangles} (black point), with coordinates
$\{12\,\text{MeV},1.32 \times 10^{−7}\}$. As we mentioned in the
previous section, that point generates a dip at $E_r=31\,$keV.  We
then generated a set of spectra by varying $m_V$ within $[1,100]\,$MeV
and $\phi$ within $[-\pi,\pi]$, for the same $|H_V|$. The results of
the $\chi^2$ analysis are displayed in the left graph in
fig. \ref{fig:chisq-analysis-NaI-Ge}, which shows the 1$\sigma$,
2$\sigma$ and 3$\sigma$ CL isocontours in the $m_V-\phi$. From this
graph it can be seen that an observation of a dip in the event
spectrum in the NaI detector cannot rule out CP violating
interactions, but can place tight bounds on $\phi$. For this
particular analysis, all values of $\phi$ but those in the range
$[-\pi/60,\pi/60]$ are excluded at the 1$\sigma$ level, and increasing
the CL does not substantially enlarge the allowed values. For the
germanium detector we use as well the point used in the previous
section (black point in the left graph in fig. \ref{fig:triangles}
located at $\{15\,\text{MeV},4.17\times 10^{-7}\}$) to generate
$N_\text{exp}$. The result of the $\chi^2$ test is shown in the right
graph in fig.~\ref{fig:chisq-analysis-NaI-Ge}. In this case, the
constraints on $\phi$ are as well competitive enough but are less
tight that those found in the NaI case. They are about a factor
$\sim 2$ less stringent due to the difference in statistics. As the
upper right and left histograms in fig. \ref{fig:dips-and-degeneracy}
show, the number of events in the NaI detector is way larger that in
the germanium one. As a consequence the statistical uncertainties in
NaI are less relevant that in Ge. Regardless of whether one includes
or not the background, which increases the statistical uncertainty,
this is a rather generic conclusion. The larger the detector the
larger the range over which $\phi$ can be excluded.
\begin{figure*}
  \centering
  \includegraphics[scale=0.373]{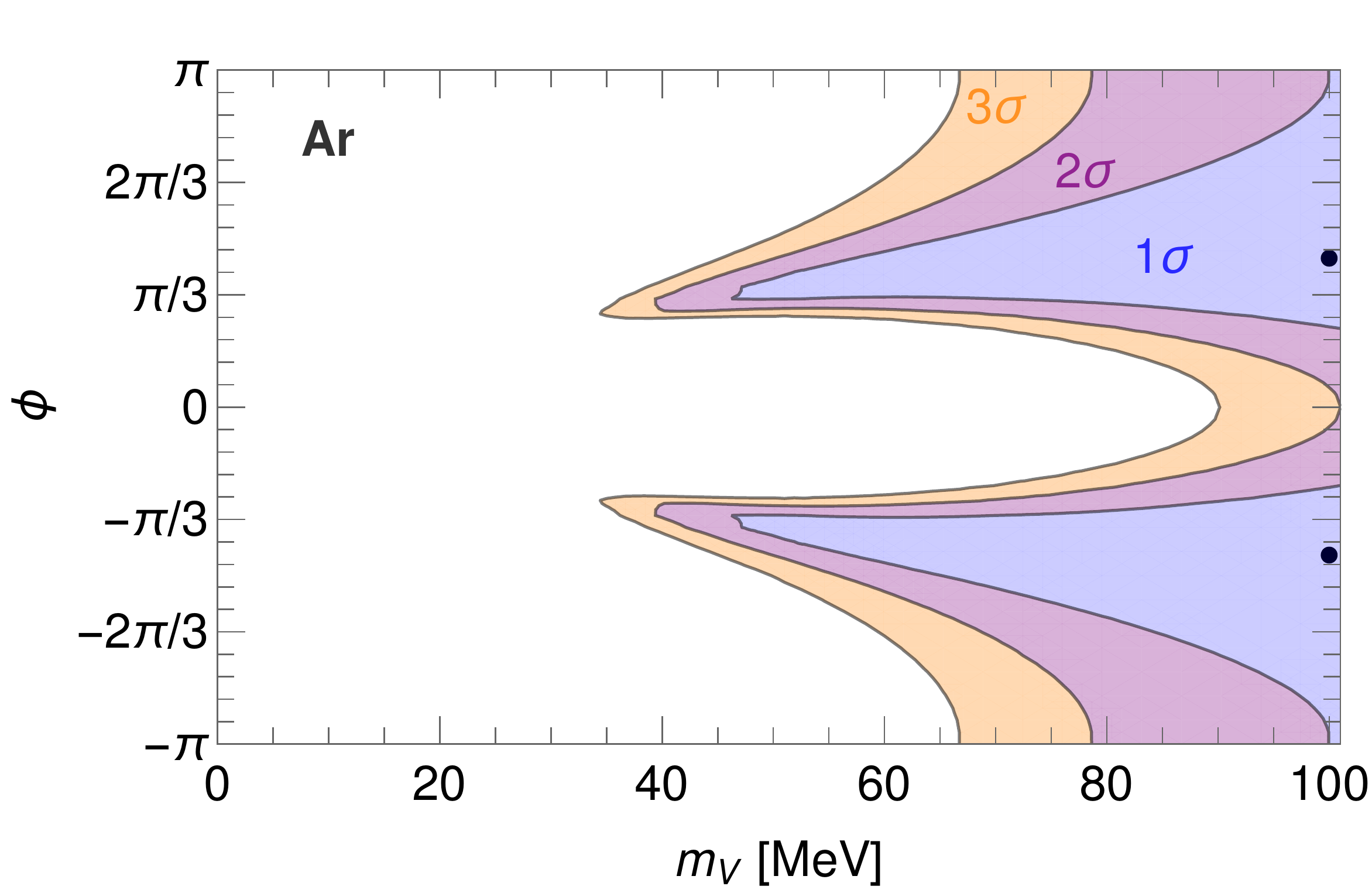}
  \hfill
  \includegraphics[scale=0.373]{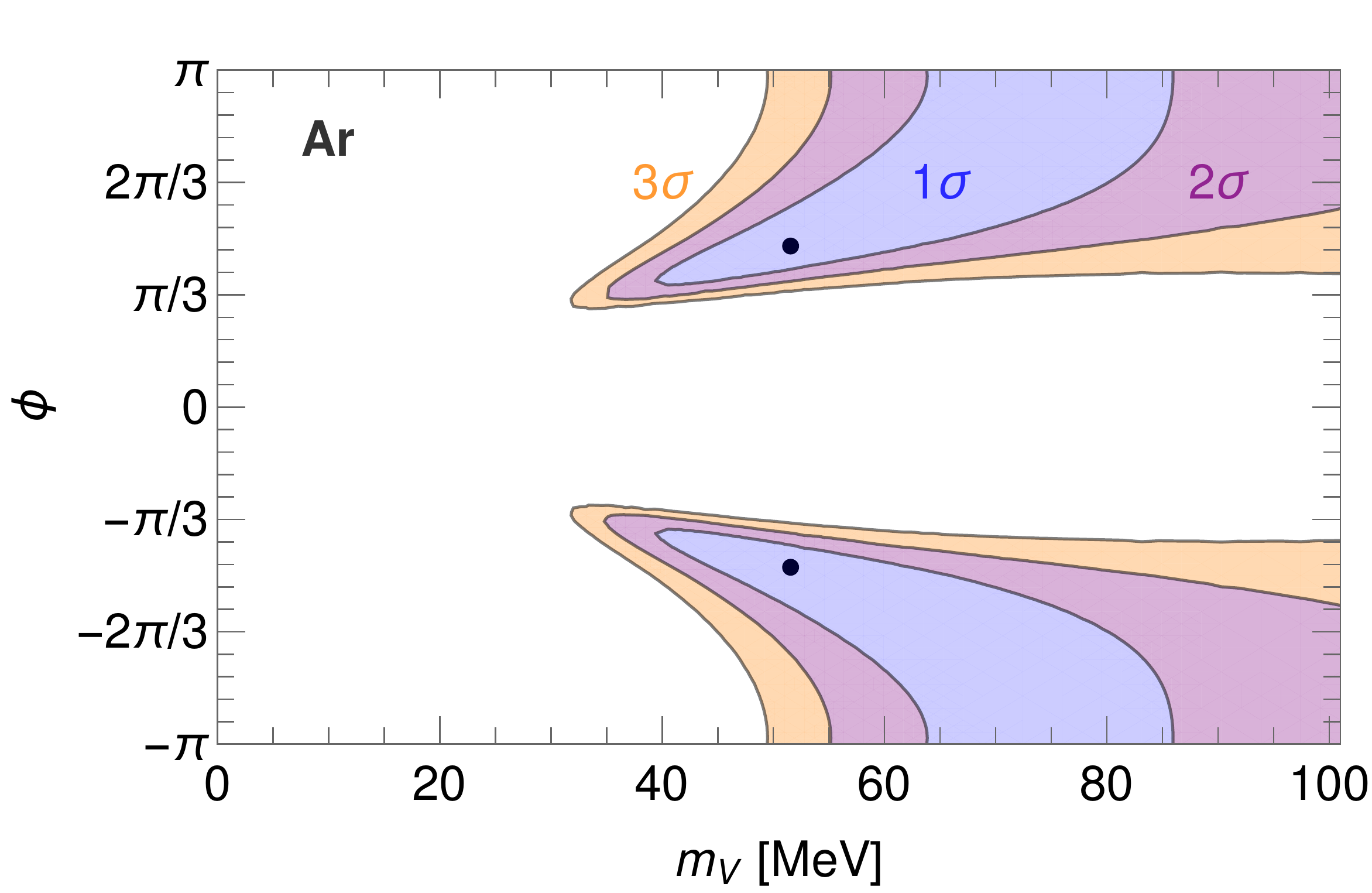}
  \caption{Results of the chi-square analysis for the LAr
    detector. \textbf{Left graph}: Regions correspond to the
    1$\sigma$, 2$\sigma$ and 3$\sigma$ CL isocontours in the
    $m_V-\phi$ plane. The contours show the regions where CP violating
    parameters (encoded in the effective CP violating phase $\phi$)
    mimic the SM event rate spectrum.  \textbf{Right graph}: Same as
    in the left graph but for degeneracy between a spectrum generated
    with real parameters and spectra generated with CP violating
    parameters. The black points indicate the best fit point values in
    both cases.}
  \label{fig:lar-constraints}
\end{figure*}
\subsection{The case of the LAr detector}
\label{sec:LAr-det}
For the LAr detector we assume the parameters shown in
tab. \ref{tab:parameters-detectors} and take for the acceptance
function a Heaviside function $H(E_r/20\text{keV}-20)$. We proceed
basically in the same way that in the sodium and germanium
detectors. For the SM degeneracy case $N_\text{exp}$ is fixed with the
SM prediction, while for the real-vs-complex degeneracy case the
pseudo-experiment data set is generated fixing $\phi$ to $\pi$,
$|H_V|$ to $4.45\times 10^{-7}$ and $m_V=16\,$MeV. For the $\chi^2$
analysis we fix $|H_V|$ to $4.25\times 10^{-7}$ and let both $m_V$
and $\phi$ vary.

The results for both analyses are shown in
fig.~\ref{fig:lar-constraints}. The left graph shows the $1\sigma$,
$2\sigma$ and $3\sigma$ CL regions for which degeneracy with the SM
prediction is induced by complex parameters. The right graph shows the
same exclusion regions for which complex parameters mimic an event
rate spectrum involving only real parameters. As we have already
stressed these results should not be understood as what the actual
experiments (or at least simulated data) will achieve, but they do
demonstrate our point: Regions in parameter space exist in which CP
violating phases can mimic signals that at first sight can be
interpreted as either SM-like or entirely generated by real
parameters. This analysis therefore allows to establish one of our
main points, that is a fully meaningful interpretation of CE$\nu$NS data in
terms of light vector mediators should come along with the inclusion
of CP violating phases.
\section{Conclusions}
\label{sec:conclusions}%
We have considered the effects of CP violating parameters on CE$\nu$NS
processes, and for that aim we have considered light vector mediator
scenarios. First of all we have introduced a parametrization that
reduces the---in principle---nine parameter problem to a three
parameter problem. We have demonstrated that this parametrization
proves to be extremely useful when dealing with CP violating
effects. In contrast to light scalar mediator schemes, light vector
mediators allow for interference between the SM and the new physics,
something that we have shown enables the splitting of the parameter space
into two non-overlapping sectors in which CP violating effects have
different manifestations: (i) A region where full destructive
interference between the SM and the new vector contribution leads to a
dip in the event rate spectrum at a certain recoil energy, (ii) a
region where CP violating parameters lead to degeneracies with either
the SM prediction or with event rate spectra generated with real
parameters.

We have shown that in case (i) information on the amount of CP
violation can be obtained. A dip in the event rate spectrum will
certainly not allow ruling out CP violation, but will allow to
place---in general---stringent constraints on the CP violating
effects, with the constraints being more pronounced with larger
detector volume. We have pointed out that the dip will as well provide
information on the real effective coupling $|H_V|$ responsible for the
signal, it will enable its reconstruction with a $4\%$ accuracy within
an interval spanning about one order of magnitude. In case (ii) we
have shown that fairly large regions in parameter space exist where CP
violating parameters can mimic CP conserving signals (SM or signals
originating from real parameters). We thus stress that meaningful and
more sensitive interpretations of future CE$\nu$NS data in terms of
light vector mediators should include CP violating parameters.

Finally, we point out that the results discussed here apply as well
for CE$\nu$NS induced by reactor or solar/atmospheric
neutrinos. Analyses of CE$\nu$NS data from these sources should
include as well CP violating effects.

\section*{Acknowledgments}
We would like to thank Danny Marfatia for reading the manuscript and
for useful comments.  To Grayson Rich for a very useful discussion on
mono-target detectors as well as for providing information regarding
different aspects of the COHERENT detectors.  We also thank Pablo
Garc\'ia and Dimitris Papoulias for useful discussions. DAS is
supported by the grant ``Unraveling new physics in the high-intensity
and high-energy frontiers'', Fondecyt No 1171136.  NR is funded by
proyecto FONDECYT Postdoctorado Nacional (2017) num. 3170135. VDR
acknowledges financial support by the ``Juan de la Cierva
Incorporaci\'on'' program (IJCI-2016-27736) funded by the Spanish
MINECO, as well as partial support by the Spanish grants
FPA2017-90566-REDC (Red Consolider MultiDark), FPA2017-85216-P and
SEV-2014-0398 (MINECO/AEI/FEDER, UE) and PROMETEO/2018/165
(Generalitat Valenciana).
\bibliography{references}
\end{document}